\documentclass[final,nonatbib]{article}

    \usepackage[final]{neurips_2020}

\usepackage[utf8]{inputenc} 
\usepackage[T1]{fontenc}    
\usepackage{hyperref}       
\usepackage{url}            
\usepackage{booktabs}       
\usepackage{amsfonts}       
\usepackage{nicefrac}       
\usepackage{microtype}      
\usepackage{xcolor}
\usepackage{graphicx,wrapfig}
\usepackage{qcircuit}
\usepackage{braket}
\usepackage{bm}

\renewcommand{\figureautorefname}{Figure~\negthinspace}

\title{Hybrid quantum-classical classifier based on tensor network and variational quantum circuit}

\author{%
   Samuel Yen-Chi Chen\\
  Computational Science Initiative\\
   Brookhaven National Laboratory\\
    Upton, NY 11973, USA \\
  \texttt{ychen@bnl.gov} \\
   \And
  Chih-Min Huang\\
Department of Physics\\
National Taiwan University\\
Taipei 10617, Taiwan\\
\texttt{b06501134@ntu.edu.tw }\\
   \And
Chia-Wei Hsing\\
Department of Physics\\
National Taiwan University\\
Taipei 10617, Taiwan\\
\texttt{cwhsing0219@gmail.com}\\
\And
Ying-Jer Kao\\
Department of Physics\\
National Taiwan University\\
Taipei 10617, Taiwan\\
\texttt{yjkao@phys.ntu.edu.tw}\\
}

\begin{document}

\maketitle

\begin{abstract}
One key step in performing quantum machine learning (QML) on  noisy intermediate-scale quantum (NISQ) devices is the dimension reduction of the input data prior to their encoding.
Traditional principle component analysis (PCA) and neural networks have been  used to perform this task; however, the classical and quantum layers are usually trained separately. 
A framework that allows for a better integration of the two key components is thus highly desirable.
Here we introduce a hybrid model combining the quantum-inspired tensor networks (TN) and the variational quantum circuits (VQC) to perform supervised learning tasks, which allows for an end-to-end training. 
We show that a matrix product state based TN with low bond dimensions performs better than PCA as a feature extractor to compress data for the input of VQCs in the binary classification of MNIST dataset.
The architecture is highly adaptable and can easily incorporate extra quantum resource when available.
\end{abstract}

\section{Introduction}
Recent growth of the quantum volume in noisy intermediate-scale quantum (NISQ) devices has stimulated rapid development in circuit-based quantum algorithms. 
In particular, quantum machine learning (QML)~\cite{schuld2018supervised,biamonte2017quantum,dunjko2018machine} using variational quantum circuits (VQC) shows great promise in surpassing the performance of classical machine learning (ML).
A VQC is a quantum circuit with adjustable parameters that are optimized according to a predefined metric, such as an objective function. 
One of the major advantages of QML compared to its classical counterpart is the drastic reduction in the number of required parameters, potentially mitigating the problem of overfitting  common in ML.
A QML architecture in modern setting typically includes a classical part and a quantum part.
Prominent examples in this hybrid genre include quantum approximate optimization algorithm (QAOA)~\cite{farhi2014quantum}, and quantum circuit learning (QCL)~\cite{mitarai2018quantum} where the VQC plays an important role as an quantum component.
Various architectures and geometries of VQC have been suggested for tasks ranging from binary classification to reinforcement learning.

One major issue in QML is how to encode classical data, typically presented in the form of high-dimensional vectors or arrays, efficiently into a quantum circuit with limited number of gate operations. 
The deep circuit depth required in either the basis or amplitude encoding makes them less desirable for the NISQ devices.  
Straightforward approaches such as  single qubit rotations promises a shallow circuit,  but suffers from the lack of representation power. 
This can be mitigated by preprocessing the input data with classical means to perform dimension reduction.
Principal component analysis (PCA) is a simple dimension reduction method and widely used in the QML research. 
More advanced methods using neural networks, though more powerful, are less commonly utilized due to the requirement of  pre-training and the significant number of parameters involved. 
Therefore, it is necessary to devise a data compression scheme which can be naturally integrated with VQC. 

In this work, we propose a hybrid framework where  a tensor network (TN)~\cite{Orus:2014um}, in particular a matrix product state (MPS)~\cite{Ostlund:1995iz,Schollwock:2011kt},  is used as a feature extractor to produce a low dimensional feature vector, which is subsequently  fed into a VQC for classification. 
Unlike other QML schemes where the classical neural network  has to be pre-trained, our framework is trained end-to-end, i.e., the MPS-VQC is trained as a whole. 
This end-to-end training indicates the quantum-classical boundary can be moved based on the available quantum resource at the training stage.
Furthermore, since the MPS can always be realized precisely by a quantum circuit~\cite{Huggins:2019kh},  the scheme is highly  adaptable and can be easily modified  when more quantum resource is provided.

The main contributions of this paper are 
\begin{itemize}
\item We propose a hybrid quantum-classical model based on TN and VQC which allows for an end-to-end training.
\item We perform a binary classification task of the MNIST dataset and show the MPS-VQC scheme is superior than a PCA-VQC scheme even at very low bond dimensions.
\item We show that the VQC serves as  regularization for the MPS to avoid over-fitting.  
\end{itemize}
%
%

\section{Methods}

\subsection{Tensor Network}
Tensor networks are efficient representation of data residing in high-dimensional space.
Originally developed to simulate quantum many-body systems,  recently TN has been applied to solve problems in classical ML~\cite{Cohen:2016mi,Stoudenmire:2016ve} and showed encouraging success in both discriminative~\cite{Levine:2018qp,Stoudenmire:2018wk,Liu:2019ty,Reyes:2020fd} and generative learning tasks~\cite{Han:2018rt}.

\begin{figure}[tbp]
\centering
\includegraphics[width=0.9\textwidth]{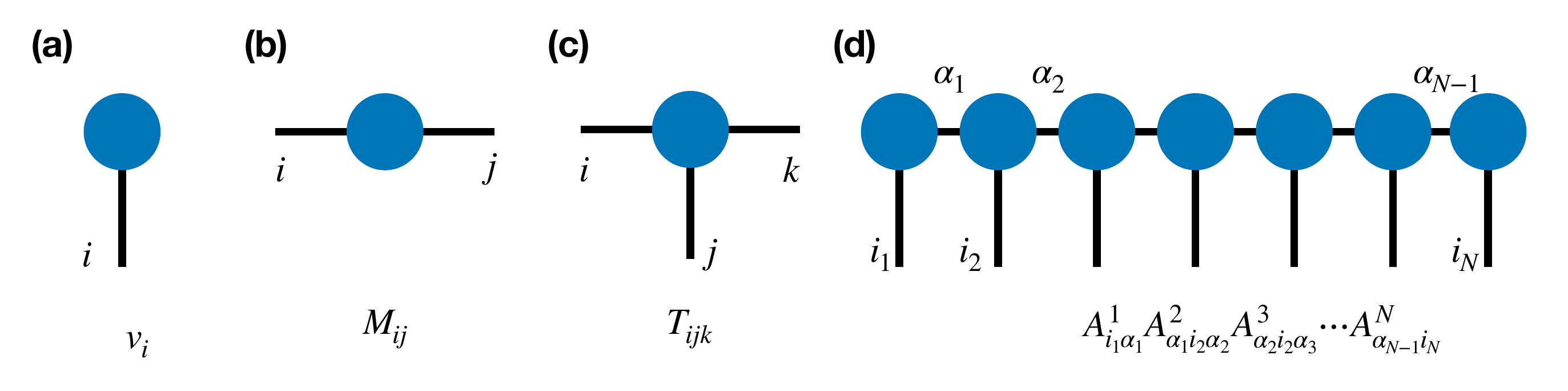}
\caption[Graphical Notation for Tensors and Tensor Networks]{{\bfseries Graphical Notation for Tensors and Tensor Networks.} (a) Graphical tensor notation for (a) a vector, (b) a matrix,  (c) a rank-3 tensor and (d) a MPS. Here we follow the Einstein convention that repeated indices, represented by internal lines in the diagram, are summed over.
}
\label{TN}
\end{figure}

It is common to use graphical notation to express tensor networks. 
A tensor is represented as a closed shape, typically a circle, with emanating lines representing tensor indices (Fig.~\ref{TN}).
The joined line indicates the corresponding index is contracted, as in the Einstein convention where repeated indices are summed over. 
The simplest TN is a MPS also known as tensor train, where tensors are contracted through the ``virtual'' indices ($\alpha$'s in Fig.~\ref{TN}(d)). 
The dimension of these virtual indices are called bond dimension and is indicated by $\chi$.
In the MPS representation of a quantum wave function, the bond dimension indicates the amount of quantum entanglement the MPS can represent in the bond. 
In the context of ML, this corresponds to the representation power of the MPS. 
The connection between the quantum entanglement and deep learning architectures has  been first explored in Ref.~\cite{Levine:2018qp,Levine:2019xt}.
In the current study, we choose the MPS as our TN for simplicity; there are other examples of TN with  distinct entanglement structures such as  the tree tensor network (TTN), multi-scale entanglement renormalization ansatz (MERA) and  projected entangled pair state (PEPS). 
The successful application of a specific TN can also give insights into the hidden correlations in the data.

The quantumness inherent in the TN gives it great advantage over other architectures in the application of QML.
In particular, since each TN can  be mapped to a quantum circuit,  it means that although in the current scheme, the TN is treated classically, it is possible to replace the whole or part of the TN component by an equivalent quantum circuit when more qubits are available.
This gives the current scheme the flexibility to move the quantum-classical boundary based on the available resources.

\subsection{Variational Quantum Circuit}
\begin{figure}[tbp]
\begin{center}
\begin{minipage}{10cm}
\Qcircuit @C=1em @R=1em {
\lstick{\ket{0}} & \multigate{3}{U(\mathbf{x})}  & \qw        & \multigate{3}{\Phi(\boldsymbol{\theta})}       & \qw      & \meter \qw \\
\lstick{\ket{0}} & \ghost{U(\mathbf{x})}         & \qw        & \ghost{\Phi(\boldsymbol{\theta})}              & \qw      & \meter \qw \\
\lstick{\ket{0}} & \ghost{U(\mathbf{x})}         & \qw        & \ghost{\Phi(\boldsymbol{\theta})}              & \qw      & \meter \qw \\
\lstick{\ket{0}} & \ghost{U(\mathbf{x})}         & \qw        & \ghost{\Phi(\boldsymbol{\theta})}              & \qw      & \meter \qw \\
}
\end{minipage}
\end{center}
\caption[Generic circuit architecture for the variational quantum classifier.]{{\bfseries Generic circuit architecture for the variational quantum classifier.}
Here $U(\mathbf{x})$ is the quantum routine for encoding classical data and $\Phi(\boldsymbol{\theta})$ is the variational circuit block with the adjustable parameters $\boldsymbol{\theta}$.
}
\label{Fig:GeneralVQC}
\end{figure}
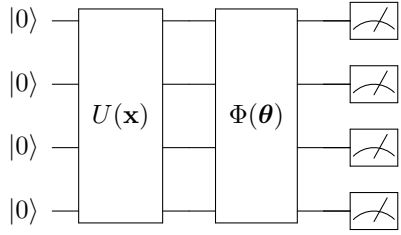

Variational quantum circuits are  quantum circuits that have {adjustable} parameters subject to classical iterative optimizations. 
The term {variational} means that certain parts of the circuit can be updated according to some predefined metric, the so-called \emph{loss}. 
We describe the general structure of a VQC in \figureautorefname{\ref{Fig:GeneralVQC}}. 
The $U(\mathbf{x})$ represents the data encoding block which is predefined and is not  optimized. 
The encoding method should be designed with respect to the problem of interest and is a crucial part in the overall architecture. 
The $\Phi(\boldsymbol{\theta})$ represents the variational block which is the \emph{learnable} part and will be optimized, usually with the gradient-based methods. 
These circuit parameters are similar to the \emph{weights} in the classical neural networks. 
It has been shown that such circuits are potentially resilient to quantum noises~\cite{kandala2017hardware,farhi2014quantum,mcclean2016theory} and therefore are suitable for building applications on NISQ devices. 

Several results have also shown that VQCs are more expressive than conventional neural networks~\cite{sim2019expressibility,lanting2014entanglement,du2018expressive} with respect to the number of parameters. 
%
Architectures based on VQCs have successfully demonstrated its capability in function approximation~\cite{mitarai2018quantum}, classification~\cite{schuld2018circuit,havlivcek2019supervised,Farhi2018ClassificationProcessors,benedetti2019parameterized}, generative modeling~\cite{dallaire2018quantum}, deep reinforcement learning~\cite{chen19} and transfer learning~\cite{mari2019transfer}.

\subsection{Hybrid TN-VQC model}

\begin{figure}[tbp]
\centering
\includegraphics[width=0.85\textwidth]{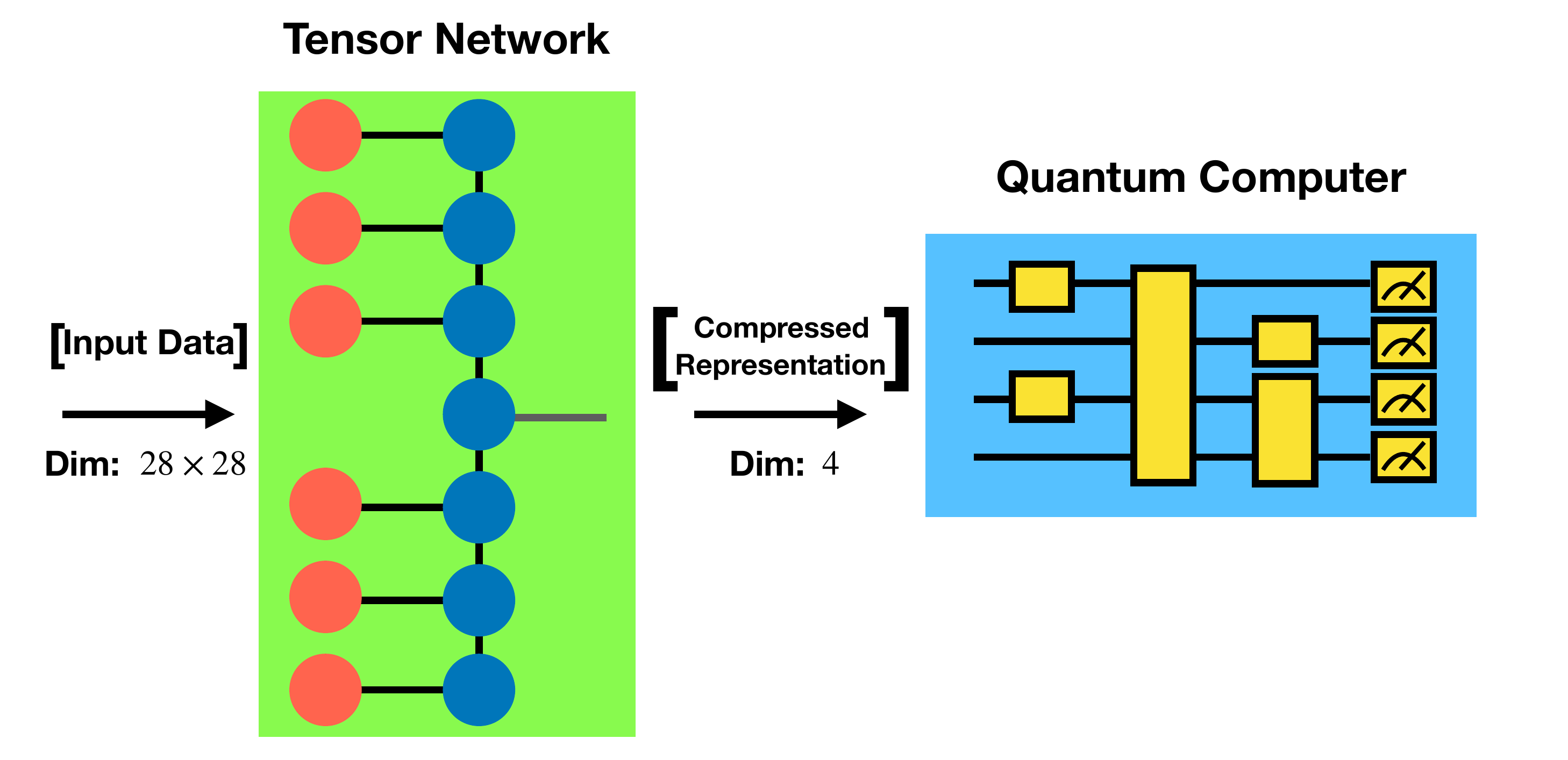}
\caption[Hybrid TN-VQC framework.]{{\bfseries Hybrid TN-VQC model.}
The TN  part is an MPS.  The number of input legs matching the dimension of input data. The output leg is a four-dimensional vector  that will be subsequently encoded into a VQC. Red circles: feature-mapped input. Blue circles: matrices with trainable parameters.}
\label{OverallArchitecture}
\end{figure}
\figureautorefname{\ref{OverallArchitecture}} shows our TN-VQC hybrid model, where the TN(MPS) serves as a feature extractor to compress the input data into a low-dimensional representation to be fed into a VQC. 
Here we follow the architecture proposed in Ref.~\cite{Stoudenmire:2016ve} for the MPS feature extractor.
Each image in the MNIST dataset is packed as an $N$-dimensional vector $\mathbf{x}=\left(x_{1}, x_{2}, \ldots, x_{N}\right)$, with $N=28\times 28=784$ and each component is normalized such that $x_i\in[0,1]$. 
The vector is mapped to a product state using the  feature map~\cite{Stoudenmire:2016ve}
\begin{equation}
\mathbf{x} \rightarrow|\Phi(\mathbf{x})\rangle=\left[\begin{array}{c}
\cos \left(\frac{\pi}{2} x_{1}\right) \\
\sin \left(\frac{x}{2} x_{1}\right)
\end{array}\right] \otimes\left[\begin{array}{c}
\cos \left(\frac{\pi}{2} x_{2}\right)\\
\sin \left(\frac{x}{2} x_{2}\right)
\end{array}\right] \otimes \cdots \otimes\left[\begin{array}{c}
\cos \left(\frac{\pi}{2} x_{N}\right) \\
\sin \left(\frac{\pi}{2} x_{N}\right)
\end{array}\right],
\end{equation}
and then fed into an MPS.
Unlike in Refs.~\cite{Stoudenmire:2016ve,Efthymiou:2019wy} where the MPS is used as a classifier, here we use the MPS as a feature extractor.
Contracting the feature-mapped input and the MPS yields a four-dimensional feature vector, which corresponds to a compressed representation to be used as an input for  the VQC to perform classification.
The dimension of the feature vector can be adjusted to fit the number of qubits available for the VQC.

\begin{figure}
\begin{center}
\begin{minipage}{10cm}
\Qcircuit @C=1em @R=1em {
\lstick{\ket{0}} & \gate{H} & \gate{R_y(\arctan(x_1))} & \gate{R_z(\arctan(x_1^2))} & \ctrl{1}   & \qw       & \qw      & \targ    & \gate{R(\alpha_1, \beta_1, \gamma_1)} & \meter \qw \\
\lstick{\ket{0}} & \gate{H} & \gate{R_y(\arctan(x_2))} & \gate{R_z(\arctan(x_2^2))} & \targ      & \ctrl{1}  & \qw      & \qw      & \gate{R(\alpha_2, \beta_2, \gamma_2)} & \meter \qw \\
\lstick{\ket{0}} & \gate{H} & \gate{R_y(\arctan(x_3))} & \gate{R_z(\arctan(x_3^2))} & \qw        & \targ     & \ctrl{1} & \qw      & \gate{R(\alpha_3, \beta_3, \gamma_3)} &  \qw \\
\lstick{\ket{0}} & \gate{H} & \gate{R_y(\arctan(x_4))} & \gate{R_z(\arctan(x_4^2))} & \qw        & \qw       & \targ    & \ctrl{-3} & \gate{R(\alpha_4, \beta_4, \gamma_4)} & \qw \gategroup{1}{5}{4}{9}{.7em}{--}\qw 
}
\end{minipage}
\end{center}
\caption[Circuit architecture for the variational quantum classifier.]{{\bfseries Variational quantum circuit architecture.}
The encoding part consists of single qubit gates and  parameters labeled $R_y(\arctan(x_i))$ and $R_z(\arctan(x_i^2))$ are for the state preparation.  
   The dashed square indicates the learnable part where the CNOT gates are used to entangle quantum states from each qubit and $R(\alpha,\beta,\gamma)$ represents the general single qubit unitary gate with three learnable parameters $\alpha_i$, $\beta_i$ and $\gamma_i$.}

\label{Fig:Basic_VQC_Hadamard}
\end{figure}
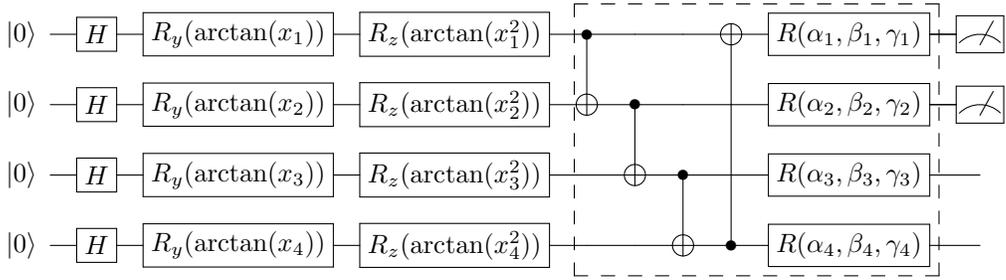

With the MPS performing dimension reduction, the rest of our hybrid framework is a VQC similar to the one proposed in Ref.~\cite{chen19}, originally designed for reinforcement learning. 
For the binary classification, we use four qubits in our VQC (Fig.~\ref{Fig:Basic_VQC_Hadamard}).
The encoding of the classical data is through  single-qubit gates $R_y(\arctan(x_i))$ and $R_z(\arctan(x_i^2))$, which represent $y$- and $z$-rotations  by the given angle $\arctan(x_i)$ and $\arctan(x_i^2)$, respectively. 
The choice of arctangent function is that in general the input values may  not be in the interval of $[-1, 1]$. 
The CNOT gates are used to entangle quantum states from each qubit and $R(\alpha,\beta,\gamma)$ represents the general single qubit unitary gate with three learnable parameters  $\alpha_i$, $\beta_i$ and $\gamma_i$. %
The first two qubits are measured for classification labels.

The current architecture can be trained end-to-end, i.e. the parameters within the MPS are updated together with those within the VQC at each iteration. 
This is a drastic contrast to other QML architectures where the classical part has to be pre-trained.
 This framework can thus be viewed as a  QML architecture where the TN part, with its own quantum circuit equivalence, is now temporarily treated classically.
%

%

\section{Experiments and Results}
To demonstrate the capabilities of the proposed TN-VQC hybrid model, we perform the binary classification task on the standard MNIST dataset, specifically digits 3 and 6. 

%

\subsection{PCA-VQC model} \label{PCA-VQC}
\begin{figure}[tbp]
\centering
\includegraphics[width=0.6\linewidth]{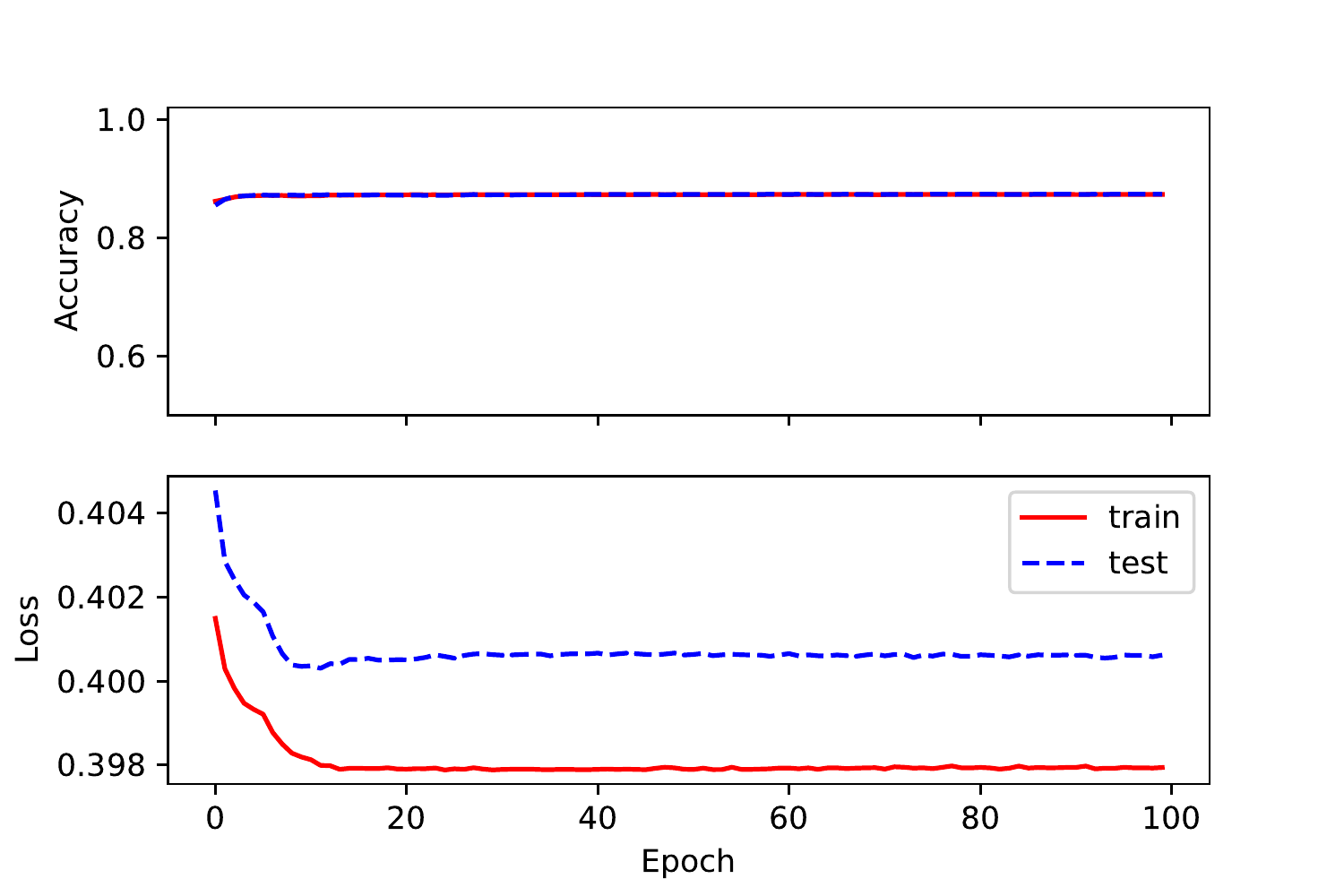}
\caption[Result: PCA preprocessing and VQC training]{{\bfseries  Result: PCA preprocessing and VQC training.}
Results of the binary classification task using PCA  to preprocess  the input data. The input data  is first reduced from dimension $28 \times 28$ to $4$ via PCA, and subsequently fed into the VQC for training.
This serves as our baseline.}
\label{EndToEnd_PCA}
\end{figure}

To develop a baseline, we study the architecture with PCA as the feature extractor and the VQC as the discriminator. 
We use PCA to reduce the input dimension of $28 \times 28 = 784$ into a four-dimensional vector, which is then fed into the VQC for training. 
In this experiment, we use RMSProp \cite{Tieleman2012} as the optimizer with the hyperparameters: learning rate $ = 0.01$, $\alpha = 0.99$ and $\epsilon = 10^{-8}$. 
The PCA is performed with the Python package scikit-learn \cite{scikit-learn}.
We can see from the result shown in~\figureautorefname{\ref{EndToEnd_PCA}} that both the accuracy and loss saturates within the first few epochs. 
As PCA has been a standard dimensionality reduction method in ML, the result validates the capability of our VQC to perform classification task.

%

\subsection{MPS classifier}
\begin{figure}[tbp]
\centering
\includegraphics[width=0.95\linewidth]{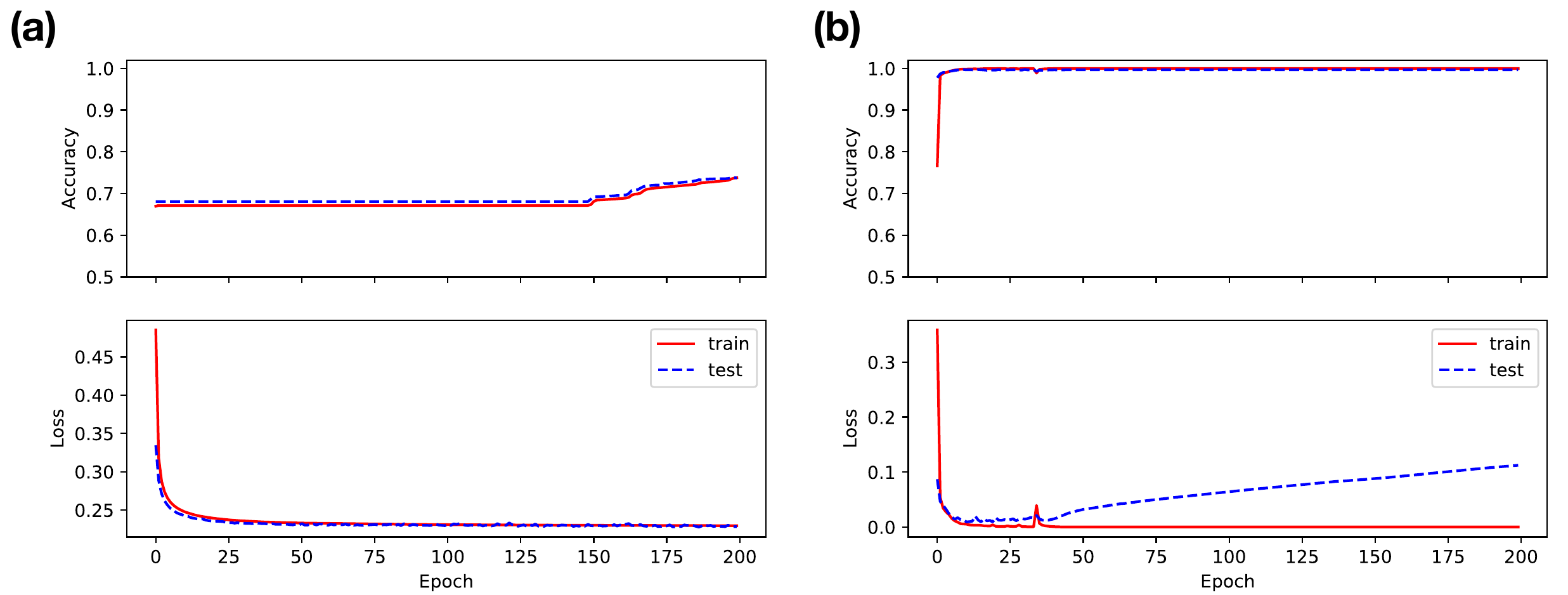}
\caption[Result: MPS classifier.]{{\bfseries Result: MPS classifier.}
Results of the binary classification task using an MPS as a classifier. Bond dimension of the MPS is (a) $\chi = 1$ (b) $\chi=2$.}
\label{MPS}
\end{figure}

Next, we demonstrate the case where the MPS part alone is fully responsible for the classification task to study its representation power. 
In this experiment, the optimizer is Adam~\cite{kingma2014adam} with a learning rate of $ 0.001$ and batch size of $100$.
Figure~\ref{MPS} shows the results of the MPS classifier for $\chi=1$ and 2. 
For $\chi=1$, the accuracy of both the training and testing datasets remain around $68-70\%$. 
When we increase the bond dimension to $\chi=2$, we observe the accuracy reaching close to $99\%$, indicating that the classifier becomes powerful enough to be highly confident in the result. 
However, we also observe that although the training loss remains low, the testing loss starts to rise. 
This implies that the classifier is getting more/less confident in the wrong/correct labels over training epochs, which could be a sign of overfitting. 
Such increasing test loss behavior is also seen in Ref.~\cite{Efthymiou:2019wy}. 

\subsection{ MPS-VQC hybrid model }
\begin{figure}[htbp]
\centering
\includegraphics[width=0.95\linewidth]{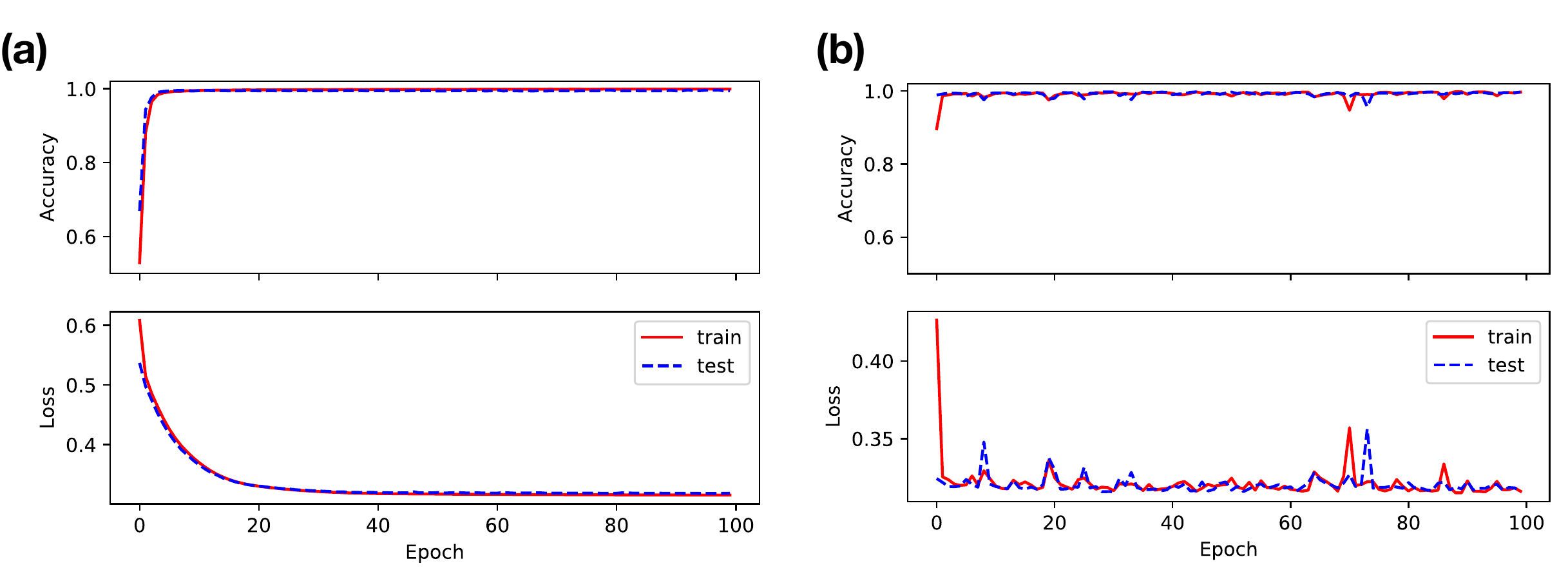}
\caption[Result: End-to-End training of MPS and VQC]{{\bfseries Result: End-to-End training of MPS-VQC model.}
In this experiment, we perform the end-to-end training of the hybrid MPS-VQC model. Bond dimension of the MPS is (a) $\chi = 1$ (b) $\chi=2$. }
\label{EndToEnd_Bond_1}
\end{figure}

Finally, we study the capability of MPS as a feature extractor for VQC. 
Here the MPS is a \emph{learnable} model with parameters subject to iterative optimization, in contrast to the PCA method presented in Section \ref{PCA-VQC}.
In this experiment, the optimizer is Adam \cite{kingma2014adam} with a learning rate of $10^{-4}$. The results are shown in~\figureautorefname{\ref{EndToEnd_Bond_1}}, where we can see that an MPS with $\chi=1$ is enough for our hybrid classifier to reach a test accuracy above $99 \%$, which is significantly better than that of the PCA-VQC model.
See \tableautorefname{\ref{tab:results_comparison}} for the performance comparison between the two methods.

For $\chi=2$ where the MPS classifier shows signs of overfitting, we find that the training of the MPS-VQC model still remains stable without the rising testing loss emerging in the MPS case, indicating that the VQC can also serve as a regularizer for the MPS part. 
We note that for $\chi>2$, the training becomes unstable and we ascribe this instability to the representation power of the model being excessive for the binary classification. Such behavior, however, still requires further investigation.

%
%

\begin{table}[tbp]
\centering
\caption{Performance comparison of  PCA-VQC and MPS-VQC ($\chi=1$).}
\label{tab:results_comparison}
\begin{tabular}{|l|l|l|l|l|}
\hline
        & Training Acc. & Testing Acc. & Training Loss & Testing Loss \\ \hline
PCA-VQC & $87.29\%$         & $87.34\%$        & $0.3979$      & $0.4006$     \\ \hline
MPS-VQC & $99.91\%$           & $99.44\%$          & $0.3154$      & $0.3183$     \\ \hline
\end{tabular}

\end{table}

\section{Discussion}
We present a hybrid quantum-classical classifier based on the quantum-inspired tensor network and the variational quantum circuit.
Such MPS-VQC framework allows for QML to deal with sizable data with limited qubits and a shallow circuit depth.
We further demonstrate the superiority of this framework by comparing it with the baseline study of a PCA-VQC model on a binary classification task of the MNIST dataset.
One clear advantage is that the representation power of the trainable MPS feature extractor is tunable with bond dimension.
It is expected that this framework can readily adapt to more difficult tasks such as the ternary classification of the Fashion-MNIST dataset. 
Our preliminary results show great promise in this direction.
\begin{ack}
This work is supported (in part) by the U.S. DOE under grant No. DE-SC-0012704 and the BNL LDRD No.20-024 and  Ministry of Science and Technology (MOST) of Taiwan under grants No. 108-2112-M-002-020-MY3 and No. 107-2112-M-002-016-MY3.
\end{ack}

\newpage

\medskip
\small
\bibliographystyle{ieeetr}
\bibliography{bib/qecc,bib/nisq,bib/vqc,bib/qml_examples,bib/qml_general,bib/machinelearning,bib/tool,M335,bib/TN}

\end{document}